\documentclass[10pt,onecolumn]{IEEEtran}
\makeatletter
\def\ps@headings{%
\def\@oddhead{\mbox{}\scriptsize\rightmark \hfil \thepage}%
\def\@evenhead{\scriptsize\thepage \hfil \leftmark\mbox{}}%
\def\@oddfoot{}%
\def\@evenfoot{}}
\makeatother
\usepackage{amssymb}
\usepackage[dvips]{graphicx}
\usepackage{mediabb}
\usepackage{cite}
\usepackage{bm}

\def\bq{\begin{equation}}
\def\eq{\end{equation}}
\def\bqn{\begin{eqnarray}}
\def\eqn{\end{eqnarray}}
\def\bqnn{\begin{eqnarray*}}
\def\eqnn{\end{eqnarray*}}

\def\bfx{{\bf x}}

\def\bfone{{\bf 1}}
\def\defeq{\stackrel{\mathrm{def}}{=}}

\newtheorem{lemma}{\bf{Lemma}}
\newtheorem{theorem}{\bf{Theorem}}
\newcommand{\triple}[1]{| #1 |_3}
\newcommand{\sizex}[1]{| #1 |_2}
\newcommand{\lengthx}[1]{| #1 |_1}
\newcommand{\anglesin}[1]{\langle \sin #1 \rangle}
\def\hatsize{\widehat{\sizex{T}}}
\def\hatlength{\widehat{\lengthx{T}}}

\title{Geometric Analysis of Observability of Target Object Shape Using Location-Unknown Distance Sensors}
\author{Hiroshi~Saito,~\IEEEmembership{Fellow,~IEEE} and Hirotada~Honda
\thanks{Manuscript received }
\thanks{Hiroshi Saito and Hirotada Honda are with NTT Network Technology Laboratories, 3-9-11, Midori-cho, Musashino-shi, Tokyo 180-8585, Japan, E-mail: saito.hiroshi@lab.ntt.co.jp, URL: http://www9.plala.or.jp/hslab/, Phone: +81 422 59 4300, Fax: +81 422 59 5671.}}

\date{}

\begin{document}
\maketitle
\begin{abstract}
We geometrically analyze the problem of estimating parameters related to the shape and size of a two-dimensional target object on the plane by using randomly distributed distance sensors whose locations are unknown.
Based on the analysis using geometric probability, we discuss the observability of these parameters: which parameters we can estimate and what conditions are required to estimate them.
For a convex target object, its size and perimeter length are observable, and other parameters are not observable.
For a general polygon target object, convexity in addition to its size and perimeter length is observable.
Parameters related to a concave vertex can be observable when some conditions are satisfied.
We also propose a method for estimating the convexity of a target object and the perimeter length of the target object.

\end{abstract}
\section{Introduction}

We investigate the problem of estimating parameters related to the shape and size of a target object by using randomly distributed distance sensors.
An individual sensor in this paper is a simple sensor measuring the distance between a sensor and a target object.
It is often composed of an infrared emitting diode and a position sensitive detector.
Its output is the voltage corresponding to the distance. 
A typical example of this sensor is a commercial sensor produced by Sharp (Sharp GP2Y0A02YK0F), although it has no communication capability. 
In this paper, we assume that an individual sensor does not have a positioning function, such as a GPS, or functions for monitoring the target object size and shape, such as a camera, and it can be placed without careful design.
However, the sensor has communication capability and sends reports.
By collecting reports from individual sensors, we statistically estimate the parameters of target objects.

The fundamental questions for this problem under this paradigm are:
Can we estimate the shape of the target object (or is its shape observable) under this paradigm?
If yes, how can we estimate it?
If no, what can we estimate? Nothing? 
This paper intends to answer these questions when distance sensors are used.

Although we also attempted to answer these questions in our prior studies \cite{infocom,signalProcess,mobileComp,time-variant}, we did not use distance sensors.
By using binary sensors of a convex sensing area reporting on whether they detect a target object, size and perimeter length estimation was proven possible even when the locations of sensors were unknown.
A size and perimeter length estimation method was proposed for use when the sensing area is convex \cite{infocom}.
To estimate additional parameters by using location-unknown sensors, combinations of binary sensors of a convex sensing area were proposed \cite{signalProcess,mobileComp}.
However, it is difficult to implement and deploy combined binary sensors called ^^ ^^ composite sensor nodes", particularly when the composite sensor nodes are large.
Another extension was done \cite{time-variant}, where the target object was assumed to be time-variant.

As far as we know, there have been no other studies to directly tackle this question.

The main contribution of this paper is the development of a concept of observability of the target object shape when location-unknown distance sensors area used.
For a convex target object, it has been proven that its size and perimeter length are observable parameters and that other parameters are non-observable.
For a general target object, convexity is also observable and conditions under which some parameters related to concave vertices become observable are provided.
We developed a method for estimating the size, perimeter length, and convexity of a target object by using distance sensors.

Unfortunately, some parameters are unobservable under the new sensing paradigm when the distance sensors are used.
This means that this paradigm has some restrictions in target object shape estimation, while it works for some applications.
For example, when we have a set of target object candidates and can distinguish them by their size, perimeter length, and convexity, this paradigm using the distance sensor can identify the target object.
This paradigm is also appropriate in other applications such as forecasting or estimating thunder areas \cite{time-variant}, because the precise estimation of the thunder area is not needed but a rough shape such as its size or length is enough.
Applications for roughly determining the environment before detailed monitoring are also suitable to this paradigm, particularly when a target object is moving.
These applications can be used in an assistant robot to detect an object of interest in a search step.

\section{Model}
There is a target object $T$ in a monitored area $\Omega\subset\mathbb{R}^2$.
$T$ is a polygon, and its boundary $\partial T$ is closed and simple (no holes or double points) and consists of $n_l$ line segments.
Let $l_i$ be the length of the $i$-th line segment and $\phi_i$ be the inner angle formed by the $i$-th and $(i+1)$-th (mod $n_l$) line segments (Fig. \ref{measured_r}-(a)), where $i=1,\cdots,n_l$.

There are $n_s$ directional distance sensors deployed in $\Omega$.
Each sensor has sensing capability to measure the distance to an object lying in the sensing direction within the maximum range $r_{max}$. 
Therefore, when the location of the sensor is $\bfx=(x,y)$ and its direction from the $x$-axis is $\theta$, the sensing range $S(\bfx,\theta)$ is $\{(x+s\cos\theta,y+s\sin\theta), 0\leq \forall s\leq r_{max}\}$, and the measured distance $r$ to $T$ by this sensor is given as 
\bq
\cases{\min_{(x+s\cos\theta,y+s\sin\theta)\in T} s, &for $S(\bfx,\theta)\cap T\neq \emptyset$,\cr \emptyset, &for $S(\bfx,\theta)\cap T=\emptyset$.}
\eq
In particular, $r=0$ if $(x,y)\in T$.
Call the point $(x+r\cos\theta,y+r\sin\theta)$ as the detected point for a given $T$, $\bfx$, and $\theta$.
Note that the detected point is on $\partial T$ if $r>0$ and in $T$ if $r=0$ and is unique.

These sensors are randomly deployed and their directions are also random and uniformly distributed in $[0,2\pi)$.
For the $i$-th sensor ($i=1,\cdots,n_s$), let $\bfx_i$ be its location, $\theta_i$ be its direction, and $r_i$ be the measured distance to $T$.
Assume that we do not know $\bfx_i$ or $\theta_i$ for any $i$.
We may remove the subscript $i$ and use $\bfx$, $\theta$, and $r$ to simplify the notation.
Because sensors monitor $\Omega$, assume that $\bfx_i\subset \Omega$ for all $i$.
To remove the boundary effect of $\Omega$, assume that the location of $T$ satisfies $\{(x',y')|\sqrt{(a-x')^2+(b-y')^2}\leq r_{max}, \forall(a,b)\subset T\}\subset \Omega$.

Each sensor can communicate with a server collecting sensing reports from individual sensors.
It reports the measured distance $r$ between the sensor and $T$ if it detects within a range or reports ^^ ^^ no detection" otherwise.
Because it does not have a positioning function or a direction sensor, the report does not include $\bfx$ or $\theta$.
All the sensors are assumed to send reports at each sensing epoch.

Although the target object and sensors can move, it is not necessary for us to assume their movement.
This is because the analysis in this paper is based on the sensing results at each sensing epoch.

For a set $X\subset \mathbb{R}^2$, $\sizex{X}$ and $\lengthx{X}$ denote its area size and perimeter length, respectively.
In addition, for $X\subset \mathbb{R}^2\times [0,2\pi)$, $\triple{X}\defeq\int\!\!\!\int\!\!\!\int_{(x,y,\theta)\in X} dx\,dy\, d\theta$.
Furthermore, $\bfone(t)\defeq\cases{1, &if $t$ is true,\cr 0, &otherwise,}$, $[t]^+\defeq \cases{t, &if $t>0$,\cr 0, &otherwise,}$,  

\noindent$\anglesin{(t)}\defeq \cases{|\sin(t)|, &if $3\pi/2<t<2\pi$,\cr 1, &if $\pi<t\leq 3\pi/2$,}$ and $\widehat{t}$ is an estimator of $t$.

Table \ref{p_list} lists the variables and parameters used in the remainder of this paper for the reader's convenience.\begin{table}
\caption{List of variables and parameters}
\begin{center}\label{p_list}
\begin{tabular}{ll}
\hline
$T$&target object\\
$\Omega$&monitored area\\
$l_i$&length of $i$-th line segment\\
$\phi_i$&inner angle formed by $i$-th and $(i+1)$-th\\
& line segments\\
$n_s$& number of sensors\\
$r_{max}$&maximum sensing range\\
$\bfx, \theta$&sensor's location and direction\\
$r$&sample value of measured distance between sensor and $T$\\
$s$&variable describing measured distance between sensor and $T$\\
$T_s(r)$&$\{(x,y,\theta)|0<\min_{(x+s\cos\theta,y+s\sin\theta)\in T}s\leq r\}$\\
$\omega(r|\theta,l)$&region where sensor lies with $s\leq r$ and \\
& direction $\theta$ and sensor's detected point is\\
& on line segment of length $l$\\
$v_i(r|\theta,l)$&overlap between $T$ and $\omega(r|\theta,l)$ for concave \\
&$\phi_i$\\
$N_0$&num. of sensing reports of which results are 0\\
$N_+$&num. of sensing reports of which results are positive\\
$N_\emptyset$&num. of sensing reports of which results are $\emptyset$\\
$N(r)$&num. of sensing reports of which results are \\
& less than $r$\\
$N(>r)$&$N_+-N(r)+N_0$\\
$N_d$&num. of sensing reports detecting $T$ ($=N_0+N_+$)\\
$e(r)$&$N(r)/n_s-\widehat{\Pr(s\leq r)}$\\
$\sigma$&(estimated) standard deviation of $e(r)$\\
\hline
\end{tabular}
\end{center}
\end{table}

\section{Integral geometry and geometric probability for our model}
As a preliminary of the analysis in this paper, this section introduces a brief explanation on integral geometry and geometric probability available for our model.

A sensing area position is characterized by the sensor location $(x,y)$ and its direction $\theta$.
We can define a probability that a set of sensing area positions satisfies a certain condition $X_c$.
An example of $X_c$ is the sensing area position detecting the target $T$.
When $(x,y)$ and $\theta$ are uniformly distributed, it is quite natural that this probability is given by the ratio of the size of the subspace $\{(x,y,\theta)|(x,y,\theta)$ satisfies $X_c\}$ to the size of the possible whole parameter domain $\Omega\times [0,2\pi)$ of $(x,y,\theta)$.
That is, the probability that a set of sensing area positions satisfies $X_c$ is $\int_{X_c}dx\,dy\,d\theta/\int_{\Omega\times [0,2\pi)}dx\,dy\,d\theta=\triple{X_c}/(2\pi\sizex{\Omega})$.
This is formally called a geometric probability based on integral geometry \cite{santalo}.
In this sense, $\int_{ X_c}dx\,dy\,d\theta$, called the measure of the set of sensor area positions satisfying condition $X_c$, is a non-normalized probability because it is proportional to the probability.

A tutorial article regarding integral geometry and geometric probability in the plane is provided at \cite{hslab} for readers who are not familiar with these subjects.

\section{Analysis}
In this section, we geometrically analyze the locations of a sensor of which the sensing result is less than $r$.
The area size of these locations gives the probability that the sensing result is less than $r$.
In this analysis, the locations and directions of the target object are fixed.

\begin{figure}[tb] 
\begin{center} 
\includegraphics[width=10cm,clip]{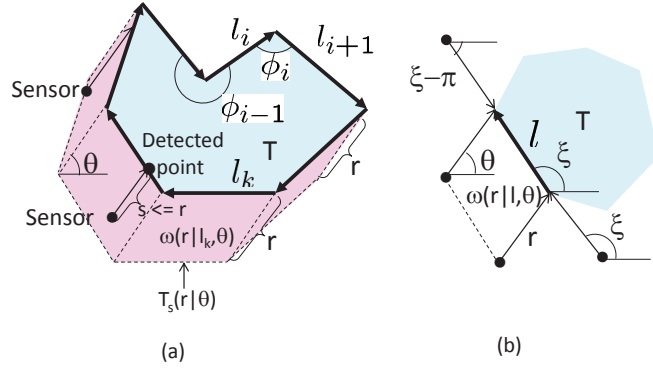} 
\caption{Illustration of locations of sensor of which sensing result is less than or equal to $r$} 
\label{measured_r} 
\end{center} 
\end{figure}

For a given direction $\theta$, the locations of a sensor of which the sensing result is less than or equal to $r$ are illustrated in Fig. \ref{measured_r}-(a).
According to the definition of geometric probability, the probability that a sensor's measured distance $s$ is smaller than or equal to $r$ is given by  
\bqn
\Pr(s\leq r)&=&\frac{\int\sizex{\tilde T_s(r|\theta)}d\theta}{\int\sizex{\Omega}d\theta}\cr
&=&(\triple{T_s(r)}+2\pi\sizex{T})/(2\pi\sizex{\Omega}),\label{pr}
\eqn
where, for a given $\theta$, 
$$\tilde T_s(r|\theta)\defeq \{(x,y)|\min_{(x+s\cos\theta,y+s\sin\theta)\in T}s\leq r\},$$
$$T_s(r)\defeq\{(x,y,\theta)|0<\min_{(x+s\cos\theta,y+s\sin\theta)\in T}s\leq r\}.$$

Let $\omega(r|\theta,l)\subset \mathbb{R}^2$ be the region where a sensor lies with $s\leq r$ and direction $\theta$, and the sensor's detected point is on the line segment of length $l$ (Fig. \ref{measured_r}-(b)).
Note that $\omega(r|\theta,l)$ is a parallelogram of edge lengths $r$ and $l$.
As shown in Fig. \ref{measured_r}-(a), $\omega(r|\theta,l)$ is a component of $T_s(r|\theta)$ where, for a given $\theta$,
$$T_s(r|\theta)\defeq \{(x,y)|0<\min_{(x+s\cos\theta,y+s\sin\theta)\in T}s\leq r\}.$$

When the direction of a line segment from the $x$-axis is $\xi$ ($0\leq \xi<2\pi$), the detected point is on this line segment if and only if $\xi-\pi\leq\theta\leq\xi$ for $r>0$ (Fig. \ref{measured_r}-(b)).
Then, for $0<r\leq r_{max}$ and fixed $\theta$,
\bq
\sizex{\omega(r|\theta,l)}=\cases{lr|\sin(\xi-\theta)|,&if $\xi-\pi\leq\theta\leq\xi$,\cr
0, &otherwise.}\label{omega_2}
\eq
Therefore, 
\bq
\triple{\omega(r|\theta,l)}=\int_{\xi-\pi}^\xi lr|\sin(\xi-\theta)|d\theta=2lr.\label{omega_3}
\eq

\subsection{Analysis for convex $T$}
For convex $T$, we derive $\Pr(s\leq r)$.
Because of the convexity of $T$, $\triple{T_s(r)}=\sum_i\triple{\omega(r|\theta,l_i)}$ (Fig. \ref{measured_r}-(a)).
Thus, according to Eq. (\ref{omega_3}),
\bq
\triple{T_s(r)}=2\sum_i l_ir=2r\lengthx{T}.\label{convex_T}
\eq
That is, $\triple{\tilde T_s(r)}=2r\lengthx{T}+2\pi\sizex{T}$.
This is fundamentally equivalent to Eq. (6.48) in \cite{santalo}.

Therefore, we obtained the following result.
\begin{theorem}
For a convex target object $T$, 
\bq
\Pr(s\leq r)=\frac{r\lengthx{T}+\pi\sizex{T}}{\pi\sizex{\Omega}}.\label{convex}
\eq
\end{theorem}

\subsection{Analysis for non-convex $T$}\label{non-convex-analysis}
\begin{figure}[tb] 
\begin{center} 
\includegraphics[width=8cm,clip]{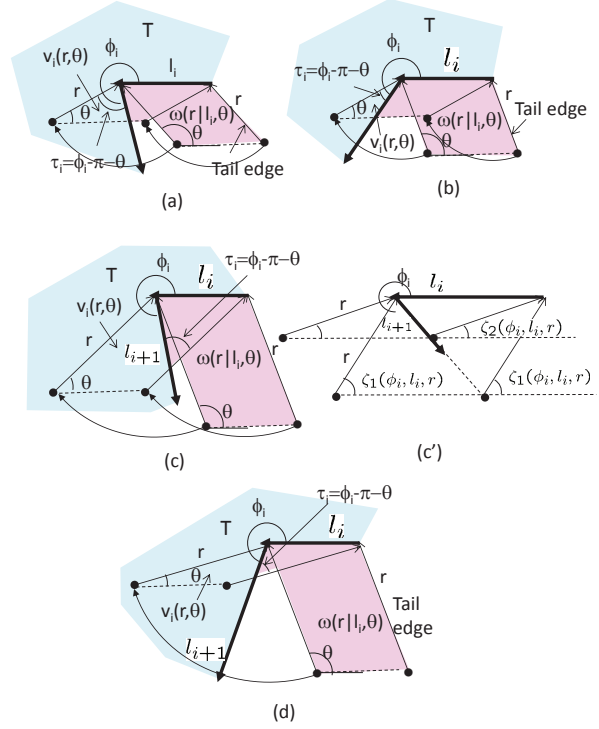} 
\caption{Illustration of cases for $\phi_i>\pi$} 
\label{non-convex} 
\end{center} 
\end{figure}

For non-convex $T$, we will derive $\triple{T_s(r)}$.
Assume that $\phi_i>\pi$, and focus on the $i$-th line segment in $\partial T$.
See Fig. \ref{non-convex} where the $x$-axis is set along this line segment to make the analysis comprehensive (this is possible because we consider this line-segment only.)
For $\phi_i>\pi$, as $\theta$ decreases, $\omega(r|\theta,l_i)$ can overlap at the concave vertex $\phi_i$ by the $(i+1)$-th line segment.
(The non-overlap part of $\omega(r|\theta,l_i)$ is in pink in Fig. \ref{non-convex}.
For a large $\theta$, the whole $\omega(r|\theta,l_i)$ is in pink.
For a small $\theta$, only a small part is in pink.)
Thus, $\triple{T_s(r)}$ may not be equal to $\sum_i\triple{\omega(r|\theta,l_i)}$.

In this paper, events in which $\omega(r|\theta,l_i)$ is limited by the $j$-th line segments within a sensing range are not taken into account, where $j\neq i-1,i+1$.

When $\theta<\phi_i-\pi$, $\omega(r|\theta,l_i)$ overlaps with $T$ (Fig. \ref{non-convex}).
Let $v_i(r|\theta,l_j)$ be the overlap between $\omega(r|\theta,l_j)$ and $T$ around $\phi_i$ where $j=i,i+1$.
In general, we need to take into account the effect of the overlaps to evaluate due to $\triple{T_s(r)}$.
Due to Eq. (\ref{omega_3}),
\bqn
&&\triple{T_s(r)}\cr
&=&\sum_i(\triple{\omega(r|\theta,l_i)}-\triple{v_i(r|\theta,l_i)}-\triple{v_i(r|\theta,l_{i+1})})\cr
&=&2r\lengthx{T} -\sum_i\sum_{l=l_i,l_{i+1}} \triple{v_i(r|\theta,l)}.\label{general}
\eqn

Consider the four exclusive cases to evaluate $\triple{v_i(r|\theta,l_i)}$ shown in Fig. \ref{non-convex} for $0<r\leq r_{max}$.
Among the edges of $\omega(r|\theta,l_i)$, consider the edge that is not parallel to the $i$-th line segment and the end point of which is at the connecting point between the $i$-th and $(i-1)$-th line segments.
We call this edge the tail edge of $\omega(r|\theta,l_i)$ because it is at the tail of the $i$-th line segment (Fig. \ref{non-convex}).
The former (latter) two cases correspond to the those in which the tail edge cannot (can) intersect the $(i+1)$-th line segment of $T$.
If the tail edge does not intersect the line segment, the overlap $v_i(r|\theta,l_i)$ is a triangle.
However, if it intersects the line segment, the overlap becomes a quadrangle.
Because the condition under which the tail edge can intersect the $(i+1)$-th line segment of $T$ depends on whether $\phi_i$ is larger than $3\pi/2$, we obtain four exclusive cases: whether $\phi>3\pi/2$ or not and whether the tail edge can intersect the $(i+1)$-th line segment or not.
Calculation of $\triple{v_i(r|\theta,l_i)}$ is fundamentally the calculation of the size of the triangle or quadrangle.
Therefore, the calculation does not require advanced mathematical knowledge, but it is messy and lengthy.
The details are shown in Appendix \ref{overlap_calculation}.

Appendix \ref{overlap_calculation} and Eq. (\ref{general}) provide us $\triple{T_s(r)}$.
\bqn
&&\triple{T_s(r)}=2\lengthx{T}r\cr
&&-\sum_{k=12,3,4}\sum_i\sum_{l=l_i,l_{i+1}}\bfone(C_{k}(\phi_i,l,r))g_k(\phi_i,l,r).\label{concave}
\eqn
That is, $\triple{v_i(r|\theta,l)}=g_k(\phi_i,l,r)$ if $C_{k}(\phi_i,l,r)$ is true where $k=12,3,4$.
Here, 
\begin{eqnarray*}
C_1(\phi_i,l_i,r)&\defeq& \{3\pi/2<\phi_i<2\pi, r<l_i|\sin\phi_i|\},\\
C_2(\phi_i,l_i,r)&\defeq& \{\pi<\phi_i\leq 3\pi/2, r<l_i\},\\
C_3(\phi_i,l_i,r)&\defeq&\{3\pi/2<\phi_i<2\pi, l_i|\sin\phi_i|\leq r\},\\
C_4(\phi_i,l_i,r)&\defeq&\{\pi<\phi_i\leq 3\pi/2, l_i\leq r\},\\
C_{12}(\phi_i,l_i,r)&\defeq& C_1(\phi_i,l_i,r)\cup C_2(\phi_i,l_i,r),\\
g_1(\phi_i)&\defeq& -2+2(\phi_i-\pi)/\tan\phi_i,\\
g_{12}(\phi_i,r)&\defeq& -r^2 g_1(\phi_i)/8,\\
\end{eqnarray*}
and $g_k(\phi_i,l,r)$ ($k=3,4$) are defined at the top of the next page.
There, $\arcsin$ takes a value between $-\pi/2$ and $\pi/2$.
\begin{figure*}[t!]
\begin{eqnarray*}
\gamma(\phi_i,l,r)&\defeq& l\sin(2\pi-\phi_i)/r,\quad
\zeta_1(\phi_i,l,r)\defeq \phi_i-\pi-\arcsin\gamma(\phi_i,l_,r),\quad
\zeta_2(\phi_i,l,r)\defeq \phi_i-\pi-(\pi-\arcsin\gamma(\phi_i,l,r)),\\
g_{3,1}(\phi_i,l,r)&\defeq& -2\gamma(\phi_i,l,r)^2+\frac{1}{\tan\phi_i}(2\arcsin\gamma(\phi_i,l,r)-2\gamma(\phi_i,l,r)\sqrt{1-\gamma(\phi_i,l,r)^2}),\\
g_{3,2}(\phi_i,l,r)&\defeq& \frac{1}{\tan\phi_i}(-2(-\arcsin\gamma(\phi_i,l,r)-\phi_i+2\pi)-2\gamma(\phi_i,l,r)\sqrt{1-\gamma(\phi_i,l,r)^2}-\sin(2\phi_i))\cr
&&-1+2\gamma(\phi_i,l,r)^2+\cos(2\phi_i),\\
g_{3,3}(\phi_i,l,r)&\defeq& \frac{l^2\sin\phi_i}{2}(\zeta_1(\phi_i,l,r)\cos\phi_i-\sin\phi_i(\log|\sin(\phi_i)|-\log(\gamma(\phi_i,l,r)))),\\
g_{3,4}(\phi_i,l,r)&\defeq& \frac{l^2\sin\phi_i}{2}(\pi-2\arcsin\gamma(\phi_i,l,r))\cos\phi_i,\\
g_3(\phi_i,l,r)&\defeq& lr(\cos([\zeta_2(\phi_i,l,r)]^+)-\cos(\zeta_1(\phi_i,l,r)))-\frac{r^2}{8}g_{3,1}(\phi_i,l,r)\\
&&+(-\frac{r^2}{8}g_{3,2}(\phi_i,l,r)+g_{3,4}(\phi_i,l,r))\bfone(\zeta_2(\phi_i,l,r)>0)+g_{3,3}(\phi_i,l,r)\bfone(\zeta_2(\phi_i,l,r)\leq 0),\\
g_4(\phi_i,l,r)&\defeq& lr(1-\cos\zeta_1(\phi_i,l,r))-\frac{r^2}{8}g_{3,1}(\phi_i,l,r)+g_{3,3}(\phi_i,l,r),
\end{eqnarray*}
\end{figure*}

The meaning of the second term of the right-hand side of Eq. (\ref{concave}) is as follows.
For a convex $\phi_i$ and $i$-th line segment (or $(i+1)$-th line segment), one of three conditions $C_{12}(\phi_i,l,r)$, $C_{3}(\phi_i,l,r)$, or $C_{4}(\phi_i,l,r)$ becomes true for a given $r$.
When $C_k(\phi_i,l,r)$ becomes true, $g_k$ provides the effect of the overlap $\triple{v_i(r|\theta,l_i)}$ (or $\triple{v_i(r|\theta,l_{i+1})}$) at $\phi_i$ and the $i$-th line segment (or $(i+1)$-th line segment).

Due to Eq. (\ref{pr}), we obtained the following result.
\begin{theorem}
For a polygon $T$ with angles $\{\phi_i\}_i$ and line segment lengths $\{l_i\}_i$,
\bqn
&&\Pr(s\leq r)\cr
&=&\frac{1}{2\pi\sizex{\Omega}}\{2\lengthx{T}r+2\pi\sizex{T}\cr
&&-\sum_{k=12,3,4}\sum_i\sum_{l=l_i,l_{i+1}}\bfone(C_{k}(\phi_i,l,r))g_k(\phi_i,l,r)\}.\label{non-convex_pr}\cr
&&
\eqn
\end{theorem}

\section{Observability}
In this paper, we use the following definition of observability and discuss what parameters of a target object we can estimate.
If some parameters are not observable, we cannot estimate them even with a huge number of sensors and with a sophisticated estimation method.

\noindent {\bf Definition of Observability}
Let $\Pr(s\leq r|\rho,\rho')$ be $\Pr(s\leq r)$ when parameter vectors $\rho$ and $\rho'$ are given.
A parameter vector $\rho$ is observable if and only if there exists a finite set of values $\{\Pr(s\leq s_i|\rho,\rho')\}_i$ that can uniquely determine $\rho$.

\subsection{Observability for convex $T$}
\begin{theorem}
When we know that $T$ is convex, $\lengthx{T}$ and $\sizex{T}$ are observable and parameters other than $\lengthx{T}$ and $\sizex{T}$ are not observable.
\end{theorem}
\proof
Because sensors are independently deployed, the sensing result of each sensor is a random sample of which probability distribution is $\Pr(s\leq r)$.
Due to Eq. (\ref{convex}), it is clear that $\lengthx{T}$ and $\sizex{T}$ are observable. 
However, parameters other than $\lengthx{T}$ and $\sizex{T}$ are not observable because Eq. (\ref{convex}) is determined only by $\lengthx{T}$ and $\sizex{T}$.
$\square$

Because $\lengthx{T}$ and $\sizex{T}$ are observable but other parameters are not observable by using randomly deployed binary sensors \cite{infocom}, this theorem implies that the parameter space that we can estimate cannot be increased even by distance sensors for convex $T$.

\subsection{Observability for general $T$}\label{non-convex-observability}
For a general polygon $T$, we will derive observable parameters and the conditions that $\{l_i,l_{i+1},\phi_i>\pi\}_i$ are observable.
\begin{theorem}
$\lengthx{T}$, $\sizex{T}$, and convexity are observable.
\end{theorem}
\proof
First, we show that $\lengthx{T}$, $\sizex{T}$, and $\sum_i\bfone(\phi_i>\pi)g_1(\phi_i)$ are observable.
Note $\Pr(s=0)=\sizex{T}/\sizex{\Omega}$.
In addition, for sufficiently small positive $s_k$ ($k=1,2$), 
$$\sum_i\sum_{l=l_i,l_{i+1}}(\bfone(C_j(\phi_i,l,s_k))=0$$
 for $j=3,4$ and  
$$\sum_i\sum_{l=l_i,l_{i+1}}\bfone(C_{12}(\phi_i,l,s_k))=2\sum_i\bfone(\phi_i>\pi).$$
Thus, $$\Pr(s\leq s_k)=\frac{\lengthx{T}s_k+\sum_i\bfone(\phi_i>\pi)s_k^2 g_1(\phi_i)/8+\pi\sizex{T}}{\pi\sizex{\Omega}},$$ where $k=1,2$.
Therefore, $\sizex{T}$, $\lengthx{T}$, and $\sum_i\bfone(\phi_i>\pi)g_1(\phi_i)$ can be determined uniquely by using $\Pr(s=0)$, $\Pr(s\leq s_1)$ and $\Pr(s\leq s_2)$ when $s_1\neq s_2$.
That is, they are observable.

Because $g_1(\phi_i)<0$ when $\phi_i>\pi$,
\bq
\cases{
\sum_i\bfone(\phi_i>\pi)g_1(\phi_i)=0& if $T$ is convex,\cr
\sum_i\bfone(\phi_i>\pi)g_1(\phi_i)<0& otherwise.
}
\eq
That is, $\sum_i\bfone(\phi_i>\pi)g_1(\phi_i)$ can represent the convexity of $T$.
Because it is observable, the convexity of $T$ is observable.
$\square$

Define 
\bqn
q(r)&\defeq& 2\pi\sizex{\Omega}\Pr(s\leq r)-2\lengthx{T}r-2\pi\sizex{T}\cr
&&+2\sum_i\bfone(\phi_i>\pi)g_{12}(\phi_i,r).\label{def_q}
\eqn
This definition implies that $q(r)$ provides an error caused by the assumption that $C_{12}(\phi_i,l,r)$ is true for any concave $\phi_i$.
Please keep in mind that $q(r)$ can be determined when $\Pr(s\leq r)$ and $r$ are given because $\lengthx{T}$, $\sizex{T}$, and  $\sum_i\bfone(\phi_i>\pi)g_1(\phi_i)$ are observable.
According to Eq. (\ref{non-convex_pr}), 
\bqn
q(r)&=&-\sum_{k=3,4}\sum_i\sum_{l=l_i,l_{i+1}}\bfone(C_{k}(\phi_i,l,r))g_k(\phi_i,l,r)\cr
&&+2\sum_i(\bfone(\phi_i>\pi)-\bfone(C_{12}(\phi_i,l,r)))g_{12}(\phi_i,r).\cr
&&\label{q_2}
\eqn
By appropriately setting $r$ in $q(r)$ and using a set of observable parameters, we can make the right-hand side of Eq. (\ref{q_2}) a function of unknown $l_i$, $l_{i+1}$, $(\phi_i,l_i)$, or $(\phi_i,l_{i+1})$ for a certain $i$ and independent of other unknown parameters.
By using this property, we can estimate $l_i$, $l_{i+1}$, $(\phi_i,l_i)$, or $(\phi_i,l_{i+1})$.
As a result, such a parameter that we can estimate becomes observable, and the set of observable parameters becomes larger.
Then, another $r$ in $q(r)$ makes 
By estimating such a parameter, the right-hand side of Eq. (\ref{q_2}) becomes a function of unknown $l_i$, $l_{i+1}$, $(\phi_i,l_i)$, or $(\phi_i,l_{i+1})$ for a another $i$.
Again, we can estimate $l_i$, $l_{i+1}$, $(\phi_i,l_i)$, or $(\phi_i,l_{i+1})$.
The conditions that makes this procedure possible is, roughly speaking, observability.

In general, parameters such as $\phi_i$ and $l_i$ are not observable.
However, $\{l_i,l_{i+1},\phi_i>\pi\}_i$ can be observable if some conditions are met.
We now provide conditions to make them observable.

\begin{theorem}
If $l_i,l_{i+1},\phi_i>\pi$ are observable, the following necessary condition for observability must be satisfied: 
\bq
r_{max}>l_i\anglesin{\phi_i},l_{i+1}\anglesin{\phi_i}.\label{necessary}
\eq
\end{theorem}
\proof
If this necessary condition is not satisfied, $\Pr(s\leq r)$ does not depend on $l_i,l_{i+1}$ for given $\lengthx{T},\sizex{T}$ because $C_{34}(\phi_i,l,s_k)\defeq C_3(\phi_i,l,s_k)\cup C_4(\phi_i,l,s_k)=0$ for $l=l_i,l_{i+1}$.
$\square$

\begin{theorem}\label{sufficient}
Assume Eq. (\ref{necessary}).
If the following sufficient condition is satisfied, $\{l_i,l_{i+1},\phi_i>\pi\}_i$ are observable:
An interval $[l_k\anglesin{\phi_i},l_k]$ does not intersect $[l_{k'}\anglesin{\phi_i},l_{k'}]$, $[l_{j}\anglesin{\phi_j},l_{j}]$, or $[l_{j+1}\anglesin{\phi_j},l_{j+1}]$ for any $j\neq i$, where $\pi<\phi_i,\phi_j$, $(k,k')=(i,i+1),(i+1,i)$.
\end{theorem}

Before the proof of this theorem, we need some preliminaries.

Under this sufficient condition, we can define the $k$-th interval $I_k$ among $\{[l_i\anglesin{\phi_i},l_i],[l_{i+1}\anglesin{\phi_i},l_{i+1}] \}_i$ in ascending order because they do not intersect each other.
That is, if $x_i\subset I_i$, $x_j\subset I_j$, and $i<j$, then $x_i<x_j$.
In addition, define $x_i\prec I_j$ as $x_i<\forall x_j\subset I_j$ and define $I_i\prec I_j$ as $x_i<x_j$ for $\forall x_i\subset I_i,\forall x_j\subset I_j$.

Let $\Xi_k$ be the set of parameters that can be uniquely determined by $\{\Pr(s\leq s_i)\}_i$, where $s_i\prec I_k$.

Roughly speaking, for $I_k\prec s'\prec I_{k+1}$, the right-hand side of Eq. (\ref{q_2}) with $r=s'$ is determined by parameters in $\Xi_k$ and an unknown parameter $l$.
Therefore, (i) determine $q(s')$ by giving $\Pr(s\leq s')$ in Eq. (\ref{def_q}), (ii) use this $q(s')$ as the left-hand side of Eq. (\ref{q_2}) with $r=s'$, and (iii) determine the unknown parameter of Eq. (\ref{q_2}) with $r=s'$.

When the right-hand side of Eq. (\ref{q_2}) with $r=s'$ is determined by parameters in $\Xi_k$ and unknown parameters $\phi,l$, introduce $s^\dagger$, where $I_k\prec s^\dagger\prec I_{k+1}$.
Note that the right-hand side of Eq. (\ref{q_2}) with $r=s^\dagger$ is also determined by parameters in $\Xi_k$ and unknown parameters $\phi,l$.
Therefore, similarly to the procedure mentioned above, by using $\Pr(s\leq s')$ and $\Pr(s\leq s^\dagger)$, we can determine the unknown parameters $\phi,l$.

By adding these parameters to $\Xi_k$, update $\Xi_k$ to $\Xi_{k+1}$ and repeat these steps.
This is the outline of the proof of Theorem \ref{sufficient}, which is given in Appendix \ref{proof_main}.

\noindent [Remark]
Equation (\ref{non-convex_pr}), which is one of the main results in this paper, does not require that the sensor locations follow a homogeneous Poisson process.
It requires that the expectation of the sensor's position is homogeneous and independent of the location of the target object.
For example, a doubly stochastic process, such as a Cox process, is possible \cite{infocom}.
The discussion regarding observability mentioned above uses this equation only.
Thus, the theoretical result of observability is valid even for a doubly stochastic process.

\section{Estimation method}
In this section, we discuss our method for estimating observable parameters of a time-invariant target object that is based on the analysis in the previous sections.
An estimation method derived in this section uses sensing reports at a single sensing epoch.
To apply this estimation method to multiple sensing epochs, we assume that sensing results at different sensing epochs are independent and are regarded as those measured by different sensors.
This assumption of independence is evaluated later in numerical examples.
When the target object or sensors move, the estimation method using sensing reports of multiple sensing epochs can reduce the number of sensors while maintaining estimation accuracy.

Although each sample of measured distance depends on the sensor location and direction, $\Pr(s\leq r)$ is of course not a function of them.
If the number of samples is large, we expect that we can observe $\Pr(s\leq r)$.
If $\Pr(s\leq r)$ is a function of a certain parameter, we expect that we can estimate it through the observed $\Pr(s\leq r)$.
Therefore, the estimation for such a parameter is possible without sensor location and direction information.

Let $N_0$, $N_+$, and $N_\emptyset$ be the number of sensing reports of which $r=0$, $r>0$, and $r=\emptyset$, respectively, where $N_0+N_++N_\emptyset=n_s$.

Because sensors are independently deployed, the likelihood for the sensing results $r_1, r_2, \cdots r_{n_s}$ is 
$$(\Pr(s=0))^{N_0}(\Pr(s=\emptyset))^{N_\emptyset}\prod_{i=N_0+1}^{n_s-N_\emptyset}\frac{d\Pr(s\leq r)}{dr}|_{r=r_i}.$$
Here, $\Pr(s=0)$ and $\Pr(s=\emptyset)=1-\Pr(s=r_{max})$ are given by $\Pr(s\leq r)$.

\subsection{Estimating parameters of convex $T$}
Due to Eq. (\ref{convex}), $\frac{d\Pr(s\leq r)}{dr}=\frac{\lengthx{T}}{\pi\sizex{\Omega}}$.
Therefore, the likelihood for the sensing results $r_1, r_2, \cdots r_{n_s}$ becomes
$(\sizex{T}/\sizex{\Omega})^{N_0}(1-\frac{r_{max}\lengthx{T}+\pi\sizex{T}}{\pi\sizex{\Omega}})^{N_\emptyset}(\frac{\lengthx{T}}{\pi\sizex{\Omega}})^{N_+}.$
Therefore, the maximum likelihood estimators $\hatlength,\hatsize$ for $\lengthx{T},\sizex{T}$ are given by the following.
\bqn
\hatlength&=&\frac{\pi N_+\sizex{\Omega}}{n_sr_{max}}\label{length}\\
\hatsize&=&N_0\sizex{\Omega}/n_s.\label{size}
\eqn
Equation (\ref{size}) is quite natural because $N_0$ is a random sample from the binomial distribution $B(n_s,\Pr(s=0))$.

These estimators are unbiased due to Eq. (\ref{convex}).
$$\cases{E[N_+]=n_s\Pr(0<s\leq r_{max})=(n_s r_{max}\lengthx{T})/(\pi\sizex{\Omega}), \cr
E[N_0]=n_s\Pr(s=0)=n_s\sizex{T}/\sizex{\Omega}.}$$

\subsection{Estimating parameters of non-convex $T$}\label{non-convex_est}
If we cannot assume that $T$ is convex, we need to use Eq. (\ref{non-convex_pr}) instead of Eq. (\ref{convex}).
However, we cannot obtain the explicit maximum likelihood estimators.

Because $\Pr(s=0)=\sizex{T}/\sizex{\Omega}$, $\sizex{T}$ can be estimated by $\Pr(s=0)$.
In addition, for sufficiently small $s_1,s_2\prec I_1$, $$\Pr(s\leq s_j)=\frac{\lengthx{T}s_j-\sum_i\bfone(\phi_i>\pi)g_{12}(\phi_i,s_j)+\pi\sizex{T}}{\pi\sizex{\Omega}},$$ where $j=1,2$.
Therefore, by using $\Pr(s\leq s_j)$ ($j=1,2$), 
\bq
\hatlength=\pi\sizex{\Omega}\frac{(s_1^2-s_2^2)\hatsize/\sizex{\Omega}+(N(s_1)s_2^2-N(s_2)s_1^2)/n_s}{s_1s_2(s_2-s_1)},\label{peri_non-convex}
\eq
and
\bq
\hatsize=N_0\sizex{\Omega}/n_s,
\eq
where $N(s_i)$ is the number of sensing results less than $s_i$.
These are unbiased estimators of $\lengthx{T}$ and $\sizex{T}$, respectively.
In particular, $\hatsize$ is given by the same formula both irrespective of the convexity of $T$.
It is also possible to estimate $\sum_i\bfone(\phi_i>\pi)g_1(\phi_i)$ by $4\sum_{j=1}^2\{\pi\sizex{\Omega}\Pr(s\leq s_i)-\lengthx{T}s_j-\pi\sizex{T}\}/s_j^2$.
However, we cannot estimate individual $g_1(\phi_i)$ or $\phi_i$ if there are multiple concave vertices.

Assume the necessary and sufficient conditions mentioned in the previous section.
Then, there exist $\{s_i\}_i$ and $\{\Pr(s\leq s_i)\}_i$ that can uniquely determine $\{l_i,l_{i+1},\phi_i>\pi\}$.
Therefore, by using $\widehat{\Pr(s\leq s_i)}$ instead of $\Pr(s\leq s_i)$, it theoretically seems possible for us to estimate these parameters.
In practice, however, it is possible only in limited situations because it is difficult to specify $\{s_i\}_i$ and because the estimation results are very sensitive to $\{s_i\}_i$, particularly when the number of concave vertices is large
(in Appendix\ref{concave_est_example}, a method for estimating $\{l_i,l_{i+1},\phi_i>\pi\}$ is provided.)

\subsection{Estimating convexity/non-convexity}
We can guess whether $T$ is convex or not with sensing results by using the following method.
(1) Assume that $T$ is convex and apply Eqs. (\ref{length}) and (\ref{size}) to obtain $\hatlength$ and $\hatsize$.
(2) Use Eq. (\ref{convex}) and replace $\lengthx{T},\sizex{T}$ with their estimates to obtain an estimate $\widehat{\Pr(s\leq r)}$.
Here, $r$ is, for example, $r_{max}/2$. 
Then, we have
\bq
\widehat{\Pr(s\leq r)}=\frac{r\hatlength+\pi\hatsize}{\pi\sizex{\Omega}}=\frac{1}{n_s}(\frac{rN_+}{r_{max}}+N_0).
\eq
(3) Define $e(r)\defeq N(r)/n_s-\widehat{\Pr(s\leq r)}$.
Note $E[e(r)]=0$ for a convex $T$ because $E[\widehat{\Pr(s\leq r)}]=E[N(r)/n_s]$ for a convex $T$.
(4) Judge that $T$ is non-convex if $e(r)$ is larger than a threshold such as $2\sigma$.  Otherwise, $T$ is convex.
Here, $\sigma$ is an estimated standard deviation of $e(r)$ and is given as follows.

Note $N_0, N(r)-N_0, N(>r)\defeq N_+-N(r)+N_0$ are samples of a multi-nomial distribution.
Therefore, 
\bqn
&&n_s^2\sigma^2\cr
&=&(\frac{r}{r_{max}})^2var[N(>r)]+(1-\frac{r}{r_{max}})^2var[N(r)-N_0]\cr
&&-2(1-\frac{r}{r_{max}})(\frac{r}{r_{max}})cov[N(>r),N(r)-N_0],
\eqn
where 
\begin{eqnarray*}
&&var[N(>r)]\cr
&=&n_s(\Pr(r<s\leq r_{max})(1-\Pr(r<s\leq r_{max}))\cr
&\approx& N(>r)(n_s-N(>r))/n_s,\cr
&&var[N(r)-N_0]\cr
&=&n_s\Pr(0<s\leq r)(1-\Pr(0<s\leq r))\cr
&\approx& (N(r)-N_0)(n_s-N(r)+N_0)/n_s,\cr
&&cov[N(>r),N(r)-N_0]\cr
&=&-n_s\Pr(r<s\leq r_{max})\Pr(0<s\leq r)\cr
&\approx& - N(>r)(N(r)-N_0)/n_s.
\end{eqnarray*}

On the basis of the judgment regarding convexity mentioned above, we propose a method for estimating $\lengthx{T}$.
The proposed method uses Eq. (\ref{length}) for $T$ judged as convex and Eq. (\ref{peri_non-convex}) for $T$ judged as non-convex.

\section{Numerical examples}
This section provides numerical examples.
With the proposed estimation method, we used parameter values $r_{max}=20$, $s_1=1$ according to the results in Appendix \ref{para-design}.
Regarding, $s_2$, we calculate $\hatlength$ using Eq. (\ref{length}) with a small number of samples and use $s_2=\hatlength/10$.

\subsection{Various $T$}\label{var_T}
We randomly generated a total of 60 polygons (20 polygons each with 4, 6, and 10 vertices) for use as a $T$.
There were 16, 4, and 4 convex polygons with 4, 6, and 10 vertices among them.

For each $T$, we conducted a simulation and applied the proposed method to judge convexity and estimate its size and perimeter length.
We conducted five simulations for each $T$.
Thus, we had a total of 300 simulation results (100 simulation results each with 4, 6, and 10 vertices).
We used $N_d=1000$ samples only to determine $s_2$.
After that, we conducted a simulation with $N_d=10000$.

First, we describe the results on the judgment of convexity/non-convexity.
For a convex $T$, the ratio of misjudgment was 0.175, 0.0667, and 0.15 for four-vertex, six-vertex, and ten-vertex target objects, respectively.
However, for a non-concave $T$, the ratio of misjudgment reached 0.55, 0.46, and 0.11.
Although the results strongly depends on the shape of $T$, it seems more difficult to find non-convexity than convexity particularly when the number of vertices is small.
To detect non-convexity, it seems that there must be a clear concave angle.
If not, we are likely to judge that $T$ is convex.
Particularly when the number of vertices is small, it seems likely that there is no clear concave vertex.
As the number of vertices increases, it is likely that there is at least one clear concave vertex.
Thus, it becomes easier to detect non-convexity.

\begin{figure}[tb] 
\begin{center} 
\includegraphics[width=8cm,clip]{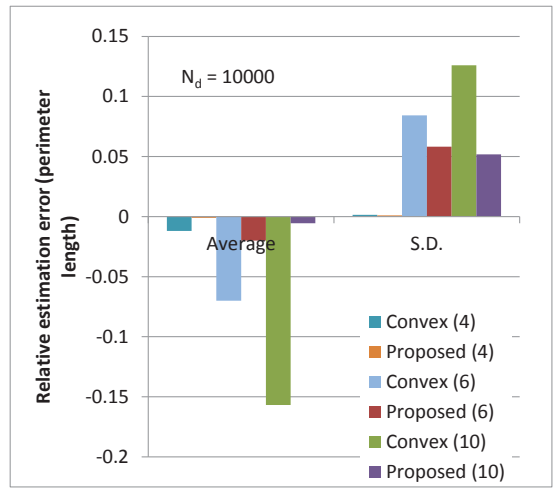} 
\caption{Estimation results of $\lengthx{T}$ for 4-, 6-, and 10-vertex target objects $T$} 
\label{peri_result} 
\end{center} 
\end{figure}

Second, we discuss the results of estimating $\lengthx{T}$.
Figure \ref{peri_result} compares the results between the proposed method and the method using Eq. (\ref{length}) for  four-, six-, and ten-vertex target objects.
As shown in this figure, both bias and variance (standard deviation) decreased for all three types (4-, 6-, and 10-vertex).

\section{Conclusion}
We investigated the estimation of parameters regarding the shape of a target object $T$ by using distance sensors of which the locations are unknown.
We analyzed the observability of parameters of $T$.
The following was shown:
When we know that $T$ is convex, the size and perimeter length are observable and other parameters are not observable.
That is, even when a huge number of sensors with a sophisticated estimation method are used, we cannot estimate parameters other than the size and perimeter length.
For a general $T$, the size, perimeter length, and convexity are observable but other parameters are, in general, not observable.
However, a concave vertex angle and the lengths of the edges forming the angle can be observable under certain conditions.

We developed a method for estimating the convexity of $T$ and the perimeter length.
The method estimating the perimeter length is applicable when we do not know whether $T$ is convex or not, because the method estimating the convexity at estimating the perimeter length.

The shape estimation using the location-unknown distance sensors have restrictions.
Not all the parameters regarding the shape can be estimated even with sophisticated data processing and a huge number of sensors.
On the other hand, this paradigm is appropriate for some applications requiring the estimation of only the size and perimeter length.

An experiment involving the proposed estimation method remains as future work.

\pagebreak

\appendices
\section{Evaluation of $\triple{v_i(r|\theta,l_i)}$}\label{overlap_calculation}

\subsection{First case}
The first case corresponds to the condition $C_1(\phi_i,l_i,r)$: $3\pi/2<\phi_i<2\pi, r<l_i|\sin\phi_i|$ (Fig. \ref{non-convex}-(a)).
In this case, the overlap occurs when $0\leq \theta \leq \phi_i-\pi$.
According to Fig. \ref{non-convex}-(a), for $\phi_i-\pi<\theta<\pi$,
\bq
\sizex{v_i(r|\theta,l_i)}=0,
\eq
and, for $0\leq \theta \leq \phi_i-\pi$,
\bq
\sizex{v_i(r|\theta,l_i)}
=\frac{r\sin\tau_i}{2}(r\cos\tau_i+\frac{r\sin\tau_i}{\tan(2\pi-\phi_i)}),
\eq
where $\tau_i\defeq \phi_i-\theta-\pi$.

\subsection{Second case}
The second case corresponds to the condition $C_2(\phi_i,l_i,r)$: $\pi<\phi_i\leq 3\pi/2, r<l_i$ (Fig. \ref{non-convex}-(b)).
According to Fig. \ref{non-convex}-(b), for $\phi_i-\pi<\theta<\pi$,
\bq
\sizex{v_i(r|\theta,l_i)}=0,
\eq
and for $0\leq \theta \leq \phi_i-\pi$,
\bq
\sizex{v_i(r|\theta,l_i)}=\frac{r\sin\tau_i}{2}(r\cos\tau_i-\frac{r\sin\tau_i}{\tan(\phi_i-\pi)}).
\eq
Because $\tan(2\pi-\phi_i)=-\tan(\phi_i)=-\tan(\phi_i-\pi)$, the equations for the second case become identical to those for the first case.
$C_{12}(\phi_i,l_i,r)$ is defined in Subsection \ref{non-convex-analysis}.

\subsection{Third case}
The third case corresponds to the condition $C_3(\phi_i,l_i,r)$: $3\pi/2<\phi_i<2\pi, l_i|\sin\phi_i|\leq r$ (Fig. \ref{non-convex}-(c)).
In this case, we need to consider three sub-cases.
When two vertices on the diagonal of the parallelogram $\omega(r|\theta,l_i)$ are on the line on which the $(i+1)$-th line segment lies, $l_i\sin(2\pi-\phi_i)=r\sin(\phi_i-\pi-\theta)$.
Let $\zeta_1(\phi_i,l_i,r),\zeta_2(\phi_i,l_i,r)$ be $\theta$ satisfying $l_i\sin(2\pi-\phi_i)=r\sin(\phi_i-\pi-\theta)$  (Fig. \ref{non-convex}-(c')).
Here, $\zeta_1(\phi_i,l_i,r)$, $\zeta_2(\phi_i,l_i,r)(\leq \zeta_1(\phi_i,l_i,r))$, and $\gamma(\phi_i,l_i,r)>0$ are defined in Subsection \ref{non-convex-analysis}, where $\arcsin$ takes a value between $-\pi/2$ and $\pi/2$.

Dependent on whether $\theta$ is larger than $\zeta_1(\phi_i,l_i,r)$ ($\zeta_2(\phi_i,l_i,r)$), the shape of $v_i(r|\theta,l_i)$ is different: triangle or quadrangle.
Therefore, we obtain the following.
For $\phi_i-\pi<\theta<\pi$,
\bq
\sizex{v_i(r|\theta,l_i)}=0,
\eq
for $\zeta_1(\phi_i,l_i,r)\leq \theta \leq \phi_i-\pi$ or for $0<\theta\leq [\zeta_2(\phi_i,l_i,r)]^+$,
\bq
\sizex{v_i(r|\theta,l_i)}
=\frac{r\sin\tau_i}{2}(r\cos\tau_i+\frac{r\sin\tau_i}{\tan(2\pi-\phi_i)}),
\eq
and for $[\zeta_2(\phi_i,l_i,r)]^+<\theta\leq \zeta_1(\phi_i,l_i,r)$,
\bqn
&&\sizex{v_i(r|\theta,l_i)}=l_ir\sin\theta-\frac{l_i\sin(2\pi-\phi_i)}{2}(l_i\cos(2\pi-\phi_i)+\frac{l_i\sin(2\pi-\phi_i)}{\tan\tau_i}).
\eqn

\subsection{Fourth case}
The fourth case corresponds to the condition $C_4(\phi_i,l_i,r)$: $\pi<\phi_i\leq 3\pi/2, l_i\leq r$ (Fig. \ref{non-convex}-(d)).
In this case, similarly to the third case, we need to consider three sub-cases.
Because $\sin(2\pi-\phi_i)=\sin(\phi-\pi)$ and $l_i\leq r$, $0<\zeta_1(\phi_i,l_i,r)<\phi_i-\pi$.
In addition, $\zeta_2(\phi_i,l_i,r)<0$.
Then, for $\phi_i-\pi<\theta<\pi$,
\bq
\sizex{v_i(r|\theta,l_i)}=0,
\eq
for $\zeta_1(\phi_i,l_i,r)\leq \theta \leq \phi_i-\pi$,
\bq
\sizex{v_i(r|\theta,l_i)}
=\frac{r\sin\tau_i}{2}(r\cos\tau_i+\frac{r\sin\tau_i}{\tan(2\pi-\phi_i)}),
\eq
and for $0<\theta\leq \zeta_1(\phi_i,l_i,r)$,
\bqn
&&\sizex{v_i(r|\theta,l_i)}=l_ir\sin\theta-\frac{l_i\sin(2\pi-\phi_i)}{2}(l_i\cos(2\pi-\phi_i)+\frac{l_i\sin(2\pi-\phi_i)}{\tan\tau_i}).
\eqn
Let $C_{34}(\phi_i,l_i,r)\defeq C_3(\phi_i,l_i,r)\cup C_4(\phi_i,l_i,r)$.

\subsection{Calculation of $\triple{v_i(r|\theta,l_i)}$}
Note that $\sizex{v_i(r|\theta,l_i)}$ is independent of the rotation of the $x$-axis, we can calculate it by assuming that the $x$-axis is set along the $i$-th line segment.
By using $\sizex{v_i(r|\theta,l_i)}$ under this assumption, we can calculate $\triple{v_i(r|\theta,l_i)}=\int\sizex{v_i(r|\theta,l_i)}d\theta$.

For each concave $\phi_i$, overlap $v_i(r|\theta,l_i)$ may occur between $T$ and $\omega(r|\theta,l_i)$.
In addition, the concave $\phi_i$ may cause the other overlap $v_i(r|\theta,l_{i+1})$ between $T$ and $\omega(r|\theta,l_{i+1})$.
Due to Eq. (\ref{general}),
\bqn
&&\triple{T_s(r)}=2r\lengthx{T}-\sum_i\sum_{j=12,3,4}\sum_{l=l_i,l_{i+1}} \bfone(C_{j}(\phi_i,l,r))\int_0^\pi\sizex{v_i(r|\theta,l)}d\theta.
\eqn
Here,
\bqn
&& \bfone(C_{12}(\phi_i,l,r))\int_0^\pi\sizex{v_i(r|\theta,l)}d\theta\cr
&=&\bfone(C_{12}(\phi_i,l,r))\{\int_0^{\phi_i-\pi}\frac{r\sin\tau_i}{2}(r\cos\tau_i-\frac{r\sin\tau_i}{\tan\phi_i})d\theta\}\cr
&=&-\bfone(C_{12}(\phi_i,l,r))r^2g_1(\phi_i)/8\cr
&=&\bfone(C_{12}(\phi_i,l,r))g_{12}(\phi_i,r)\label{appendix_g1}\cr
&&
\eqn
where $g_1(\phi_i), g_{12}(\phi_i,r)$ are defined in Section \ref{non-convex-analysis}.
This is because
\bqn
&&\int_a^b \frac{r\sin\tau_i}{2}(r\cos\tau_i-\frac{r\sin\tau_i}{\tan\phi_i})d\theta\cr
&=&[-\frac{r^2}{8}(-\cos(2\tau_i)+\frac{-2\tau_i+\sin(2\tau_i)}{\tan\phi_i})]_{\tau_i=\phi_i-a-\pi}^{\phi_i-b-\pi}.\label{messy}\cr
&&
\eqn
In addition,
\bqn
&&\bfone(C_{3}(\phi_i,l,r))\int_0^\pi\sizex{v_i(r|\theta,l)}d\theta\cr
&=&\bfone(C_{3}(\phi_i,l,r))\{\int_{\zeta_1(\phi_i,l,r)}^{\phi_i-\pi}\frac{r\sin\tau_i}{2}(r\cos\tau_i-\frac{r\sin\tau_i}{\tan\phi_i})d\theta\cr
&&+\int_0^{[\zeta_2(\phi_i,l,r)]^+}\frac{r\sin\tau_i}{2}(r\cos\tau_i-\frac{r\sin\tau_i}{\tan\phi_i})d\theta\cr
&&+\int_{[\zeta_2(\phi_i,l,r)]^+}^{\zeta_1(\phi_i,l,r)} lr\sin\theta+\frac{l\sin\phi_i}{2}(l\cos\phi_i-\frac{l\sin\phi_i}{\tan\tau_i})d\theta\},\cr
&=&\bfone(C_{3}(\phi_i,l,r))g_3(\phi_i,l,r),\cr
&&
\eqn
where $g_{3,1}(\phi_i,l,r)$, $g_{3,2}(\phi_i,l,r)$,  $g_{3,3}(\phi_i,l,r)$, $g_{3,4}(\phi_i,l,r)$ and $g_3(\phi_i,l,r)$ are defined in Secction \ref{non-convex-analysis}.
This is due to Eq. (\ref{messy}) and because, for $j=1,2$,
\bqn
&&\cos(2(\phi_i-\pi-\zeta_j(\phi_i,l_i,r)))=1-2\gamma(\phi_i,l_i,r)^2,\\
&&\sin(2(\phi_i-\pi-\zeta_j(\phi_i,l_i,r)))\cr
&&=\cases{2\gamma(\phi_i,l_i,r)\sqrt{1-\gamma(\phi_i,l_i,r)^2}, &for $j=1$,\cr
-2\gamma(\phi_i,l_i,r)\sqrt{1-\gamma(\phi_i,l_i,r)^2}, &for $j=2$.}
\eqn
Furthermore,
\bqn
&&\bfone(C_{4}(\phi_i,l,r))\int_0^\pi\sizex{v_i(r|\theta,l)}d\theta\cr
&=&\bfone(C_{4}(\phi_i,l,r))\{\int_{\zeta_1(\phi_i,l,r)}^{\phi_i-\pi}\frac{r\sin\tau_i}{2}(r\cos\tau_i-\frac{r\sin\tau_i}{\tan\phi_i})d\theta\cr
&&+\int_0^{\zeta_1(\phi_i,l,r)} lr\sin\theta+\frac{l\sin\phi_i}{2}(l\cos\phi_i-\frac{l\sin\phi_i}{\tan\tau_i})d\theta\},\cr
&=&\bfone(C_{4}(\phi_i,l,r))g_4(\phi_i,l,r).\cr
&&
\eqn
where $g_4(\phi_i,l,r)$ is defined in Section \ref{non-convex-analysis}.

\section{Proof of Theorem \ref{sufficient}}\label{proof_main}
Note that $I_k$ has the following characteristic.
\begin{lemma}\label{lem1}
Assume the sufficient condition in Theorem \ref{sufficient}.
When $s^\dagger\prec I_k \prec s'\prec I_{k+1}$, there is only a single pair $(\phi,l)\in\{(\phi_i,l_i),(\phi_i,l_{i+1})\}_i$ such that 
\bqnn
\bfone(C_{12}(\phi,l,s^\dagger))&=&1,\cr
\bfone(C_{34}(\phi,l,s^\dagger))&=&0,\cr
\bfone(C_{12}(\phi,l,s'))&=&0,\cr
\bfone(C_{34}(\phi,l,s'))&=&1.
\eqnn
\end{lemma}
This lemma is clear due to the definitions of $C_{12}$, $C_{34}$, and $\{I_k\}_k$.

\begin{lemma}
When $r=l+\epsilon$,
\bq
\sum_{k=3,4}\bfone(C_{k}(\phi,l,r))g_k(\phi,l,r)\
=\bfone(C_{34}(\phi,l,r))g_4(\phi,l,r) \label{epsilon}
\eq
where $\epsilon$ is a very small positive constant.
\end{lemma}
\proof
When $\bfone(C_3(\phi,l,l+\epsilon))=0$ and $\bfone(C_4(\phi,l,l+\epsilon))=1$, it is clear that Eq. (\ref{epsilon}) is valid.

Assume that $\bfone(C_3(\phi,l,l+\epsilon))=1$ and $\bfone(C_4(\phi,l,l+\epsilon))=0$.
This means $3\pi/2<\phi<2\pi$.
Then, $\arcsin \gamma(\phi,l,l+\epsilon)=2\pi-\phi-\epsilon'$ where $\epsilon'$ is a very small positive constant.
Thus, $\zeta_2(\phi,l,l+\epsilon)=-\epsilon'<0$.
When $\zeta_2(\phi,l,l+\epsilon)<0$, $g_3(\phi,l,l+\epsilon)=g_4(\phi,l,l+\epsilon)$.
Therefore, Eq. (\ref{epsilon}) is valid.
$\square$

\subsection{Main part of proof of Theorem \ref{sufficient}}
When $\phi_i,l_i\in \Xi_k$ and $l_{i+1}\not\in\Xi_k$ ($\phi_i,l_{i+1}\in \Xi_k$ and $l_{i}\not\in\Xi_k$), we call $l_{i+1}$ ($l_i$) the undetermined line segment of $(\phi_i,\Xi_k)$ and use the notation $l(\phi_i,\Xi_k)$ to express this undetermined line segment.
When $I_k=[l_i\anglesin{\phi_i},l_i]$, define $l(I_k)=l_i$.
Set $\Xi_1=\emptyset$.

For $s'\prec I_1$, $\sum_{l=l_j,l_{j+1}}\bfone(C_{34}(\phi_j,l,s'))=0$ for any $j$. 
Therefore, due to Lemma \ref{lem1}, for $I_1\prec s_3<s_4\prec I_2$, there exists a single pair $(\phi_i,l)$ that satisfies the following conditions with $k=3,4$ where $l=l_i$ or $l_{i+1}$:
$$\cases{\bfone(C_{34}(\phi_i,l,s_k))>0,\cr \bfone(C_{34}(\phi_j,l,s_k))=0,&for any $j\neq i$. }$$

We can assume $\bfone(C_{34}(\phi_i,l_i,s_k))=1,\bfone(C_{34}(\phi_i,l_{i+1},s_k))=0$ without loss of generality.
Set $s_k-l(I_1)=\epsilon_k$, a very small positive constant, for $k=3,4$, and apply Eq. (\ref{epsilon}) with $l=l(I_1),r=l(I_1)+\epsilon_k$.
Then, due to Eq. (\ref{q_2}), we can obtain $$q(s_k)=-g_4(\phi_i,l_i,s_k)+g_{12}(\phi_i,s_k)$$ ($k=3,4$).
By using a given $q(s_3)$ and $q(s_4)$, unknown parameters $\phi_i$ and $l_i$ are uniquely determined (see Appendix \ref{unique_2p}).
Update $\Xi_0$ by adding the parameters determined here.
That is, set $\Xi_1=\Xi_0\cup\{\phi_i,l_i\}=\{\phi_i,l_i\}$.

Similarly, for $I_2\prec r\prec I_3$, we can determine a pair $(\phi,l)\not\in\Xi_1$ or a undetermined line segment $l(\phi,\Xi_1)$.
Note, for $I_2\prec r\prec I_3$, (1) or (2) defined below occurs:

\noindent (1) there exists a single pair $(\phi,l)\not\in\Xi_1$ that $$\cases{\sum_{l}\bfone(C_{34}(\phi,l,r))>0, \cr \sum_{l}\bfone(C_{34}(\phi',l,r))=0,& for any $\phi'\not\in\Xi_1\cup \{\phi\}$,}$$

\noindent (2) there exists $(\phi,l)\in\Xi_1$ that $$\cases{\bfone(C_{34}(\phi,l,r))=1,\cr \bfone(C_{34}(\phi,l(\phi,\Xi_1),r))=1,\cr \sum_{l=l_j,l_{j+1}}\bfone(C_{34}(\phi_j,l,r))=0,& for any $\phi_j\not\in\Xi_1$.}$$

For (1), we can assume $\bfone(C_{34}(\phi_k,l_k,r))=1$, $\bfone(C_{34}(\phi_k,l_{k+1},r))=0$ without loss of generality.
Then, for $I_2\prec s_5<s_6\prec I_3$, set $s_j-l(I_2)$ to be a very small positive constant $\epsilon_j$ for $j=5,6$, and apply Eq. (\ref{epsilon}) with $l=l(I_2),r=l(I_2)+\epsilon_j$.
Then, due to Eq. (\ref{q_2}), 
\bqnn
&&q(s_j)+ \sum_{m=3}^4\sum_{\phi,l\in\Xi_1}\bfone(C_{m}(\phi,l,s_j))g_m(\phi,l,s_j)-g_{12}(\phi,s_j)=-g_4(\phi_k,l_k,s_j)+g_{12}(\phi_k,s_j),
\eqnn
where $j=5,6$.
Because we have already uniquely determined $\phi_i,l_i$, the left-hand side of the above equation is given.
Thus, $\phi_k,l_k$ are uniquely determined (see Appendix \ref{unique_2p}).
Update $\Xi_1$ by adding the parameters determined here.
That is, set $\Xi_2=\Xi_1\cup\{\phi_k,l_k\}$.

For (2), for $I_2\prec s_5\prec I_3$, set $s_5-l(I_2)$ to be a very small positive constant $\epsilon_5$ and apply Eq. (\ref{epsilon}) with $l=l(I_2),r=l(I_2)+\epsilon_5$.
Then, due to Eq. (\ref{q_2}), 
\bqnn
&&q(s_5)+ \sum_{m=3}^4\sum_{\phi,l\in\Xi_1}\bfone(C_{m}(\phi,l,s_5))g_m(\phi,l,s_5)-2g_{12}(\phi,s_5)=-g_4(\phi_i,l(\phi_i,\Xi_1),s_5).
\eqnn
By using the given left-hand side of the above equation, $l(\phi_i,\Xi_1)$ is uniquely determined (see Appendix \ref{unique_1p}).
Update $\Xi_1$ by adding the parameters determined here.
That is, set $\Xi_2=\Xi_1\cup\{l(\phi_i,\Xi_1)\}$.

By repeating this procedure, we can uniquely determine $\lengthx{T}$, $\sizex{T}$, and $\{l_i,l_{i+1},\phi_i>\pi\}_i$.



\section{Proof of uniqueness}
\subsection{Proof of uniqueness regarding $\phi$ and $l$}\label{unique_2p}
In this section, we use the notations 
\bqn
&& \hspace{5mm} {\cal G}_1 \equiv \Bigl\{ (\phi,l,r) | \pi < \phi \leq 3\pi/2, \; l \leq r \Bigr\}, \nonumber \\
&& \hspace{0mm} {\cal G}_2 \equiv \Bigl\{ (\phi,l,r) | 3\pi/2 < \phi < 2\pi, \; l|\sin\phi|\leq r \Bigr\}. \nonumber
\eqn
Below, we define $\arcsin(\cdot)$ with its value on the interval $(-\pi/2,\pi/2)$ and
$\arccos(\cdot)$ on $(0,\pi)$. 
We also simplify the representations of some of the functions below. They were obtained with the aid of elementary calculations,
and we omit the proof:
\bqn
\left\{
\begin{array}{l}\label{39}
 \displaystyle \cos (\zeta_1(l,\phi,r)) = - \sqrt{1- \bigl( \gamma(\phi,l,r) \bigr)^2} \cos \phi \\[5pt]
 \displaystyle  \hspace{25mm}+ lr^{-1} \sin^2 \phi,  \\[12pt]
 \displaystyle g_{3,1}(l,\phi,r) = -2 \Bigl( \frac{l\sin\phi}{r} \Bigr)^2 + \frac{2\arcsin \gamma(l,\phi,r)}{\tan \phi} \\[5pt]
 \displaystyle \hspace{25mm}
 \displaystyle  + 2lr^{-1} \sqrt{1-\bigl( \gamma(\phi,l;r) \bigr)^2} \cos \phi, \\[12pt]
 \displaystyle \hspace{13mm}
 \displaystyle g_{1}(\phi) = -2 \Biggl\{ 1- \frac{(\phi-\pi)}{\tan \phi} \Biggr\}. 
\end{array}
\right.
\eqn
We prove the uniqueness of a solution when the observability condition holds.

Using Eq. (\ref{39}), we obtain the following equality.
\bqn\label{40}
\Phi(\phi,l,r) &\equiv& -g_4(\phi,l,r)-\frac{r^2}{8}g_1(\phi)  \nonumber \\[5pt]
 &=&-lr + \frac{r^2}{4} -\frac{3l}{4} \sqrt{r^2 - (l\sin \phi)^2} \cos \phi \nonumber \\[5pt]
 &&\hspace{0mm} 
  + \frac{3}{4}\bigl( l\sin\phi\bigr)^2 - \frac{r^2\zeta_1(\phi,l,r)}{4\tan \phi} \nonumber \\[5pt]
 &&\hspace{0mm}
  -\frac{l^2 \zeta_1(\phi,l,r)}{2} \sin\phi \cos\phi  \nonumber \\[5pt]
 &&\hspace{0mm}
  + \Bigl( \log(r/l) \Bigr) \frac{(l\sin\phi)^2}{2}
\eqn

Now, we introduce the notation $\boldsymbol{x} = (\phi,l)$ and regard $r$ as a parameter for the meanwhile.
We also use the notation $\Phi(\boldsymbol{x};r) \equiv \Phi(l,\phi,r)$ hereafter.
Then, we prove uniqueness in accordance with the following argument.

Let us suppose that two points $\{\boldsymbol{x}_j \}_{j=1,2} = \{(\phi_j,l_j)\}_{j=1}^2$ satisfy 
$$
\Phi(\boldsymbol{x}_1;r_i)=\Phi(\boldsymbol{x}_2;r_i)=d_i \; (i=1,2).
$$

Since this means
$$
\Phi(\boldsymbol{x}_1;r_i) - \Phi(\boldsymbol{x}_2;r_i) = 0 \; (i=1,2),
$$
by virtue of the mean value theorem, we have
\begin{eqnarray}\label{41}
\displaystyle
 \bigl( \boldsymbol{x}_1 -\boldsymbol{x}_2 \bigr) \cdot 
  \int_0^1 \nabla \Phi(\eta \boldsymbol{x}_1
   + (1-\eta) \boldsymbol{x}_2;r_i) \; {\rm d}\eta =0 \nonumber \\
\displaystyle (i=1,2),
\end{eqnarray}
where $\nabla = (\partial / \partial \phi,  \partial / \partial l)^{\rm T}$.
Let us introduce
\begin{eqnarray*}
&\hspace{0mm}\displaystyle
 \tilde{\boldsymbol{x}} = \boldsymbol{x}_1 -\boldsymbol{x}_2, \hspace{10mm} \\
&\displaystyle
 f_1(r;\boldsymbol{x}_1,\boldsymbol{x}_2)
 \equiv \int_0^1 
  \frac{\partial \Phi}{\partial \phi} (\eta \boldsymbol{x}_1 + (1-\eta) \boldsymbol{x}_2;r) \; {\rm d}\eta,\\
&\displaystyle
 f_2(r;\boldsymbol{x}_1,\boldsymbol{x}_2)
 \equiv 
  \int_0^1 \frac{\partial \Phi}{\partial l} (\eta \boldsymbol{x}_1 + (1-\eta) \boldsymbol{x}_2;r) \; {\rm d}\eta.
\end{eqnarray*}
Then, Eq. (\ref{41}) equals 
\begin{eqnarray}
\boldsymbol{M}(\boldsymbol{x}_1,\boldsymbol{x}_2) \tilde{\boldsymbol{x}} = {\bf 0},\nonumber
\end{eqnarray}
where 
$$
\boldsymbol{M}(\boldsymbol{x}_1,\boldsymbol{x}_2) = [m_{ij}]_{i,j=1,2}
$$
with 
$m_{i,j} = f_j(r_i;\boldsymbol{x}_1,\boldsymbol{x}_2)$. Therefore, if 
\begin{eqnarray}
\det \boldsymbol{M} (\boldsymbol{x}_1,\boldsymbol{x}_2) \ne 0\label{42}
\end{eqnarray}
 holds for every 
$(\boldsymbol{x}_1,\boldsymbol{x}_2),$ it means $\tilde{\boldsymbol{x}} = {\bf 0}$, which directly leads to the desired statement.

Now, assume $r_1 < r_2$ without loss of generality; then, Eq. (\ref{42}) amounts to
\begin{eqnarray}\label{43}
&f_1(r_1;\boldsymbol{x}_1,\boldsymbol{x}_2)f_2(r_2;\boldsymbol{x}_1,\boldsymbol{x}_2) \hspace{30mm} \nonumber \\
& \quad - f_1(r_2;\boldsymbol{x}_1,\boldsymbol{x}_2)f_2(r_1;\boldsymbol{x}_1,\boldsymbol{x}_2) \ne 0.
\end{eqnarray}
Thereby, it is sufficient to show, for instance,
\begin{eqnarray*}
f_1(r_1;\boldsymbol{x}_1,\boldsymbol{x}_2) \geq f_1(r_2;\boldsymbol{x}_1,\boldsymbol{x}_2), \; \\
f_2(r_1;\boldsymbol{x}_1,\boldsymbol{x}_2) \leq f_1(r_2;\boldsymbol{x}_1,\boldsymbol{x}_2) \quad 
\forall \boldsymbol{x}_1,\boldsymbol{x}_2
\end{eqnarray*}
for Eq. (\ref{43}). In particular, it suffices to show 
\bqn
&&\frac{\partial \Phi}{\partial l}(\phi,l,r_1)
 \geq \frac{\partial \Phi}{\partial l}(\phi,l,r_2), \label{44} \\[5pt]
&&\frac{\partial \Phi}{\partial \phi}(\phi,l,r_1)
 \leq \frac{\partial \Phi}{\partial \phi}(\phi,l,r_2)  \label{45}\\[5pt]
&&\hspace{10mm}
 \forall (\phi, l, r_i ) \in \bigcup_{i=1}^2 {\cal G}_i, \; r_1<r_2, \nonumber
\eqn
and we shall prove Eqs. (\ref{44})--(\ref{45}) below.

It is obvious that $\Phi(\phi,l,r) $ is smooth enough with respect to 
$(\phi,l,r)$, and we first derive $\partial \Phi/\partial r$.
By virtue of Eq. (\ref{40}), after some calculations, we have
\bqn
&&\displaystyle
 \frac{\partial \Phi}{\partial r} = -l+\frac{r}{2} -\frac{l\cos \phi}{2r}\sqrt{r^2-(l\sin\phi)^2}
 + \frac{(l\sin\phi)^2}{2r}  \nonumber \\[2pt]
&&\hspace{10mm}
 \displaystyle
 -\frac{r}{2\tan \phi}\Bigl\{ \phi-\pi + \arcsin (l\sin\phi/r) \Bigr\}.\label{46}
\eqn
From this, we deduce
\begin{eqnarray}
&\displaystyle
 \frac{\partial^2 \Phi}{\partial l \partial r} 
 = \frac{1}{r} \Bigl(
                -r + l\sin^2\phi -\sqrt{ r^2-(l\sin\phi)^2 }  \cos \phi \label{47}
               \Bigr).
\end{eqnarray}
Since $-r+l\sin^2\phi \leq 0$, it is sufficient to consider the case $\cos\phi \leq 0$ to show the non-positiveness of 
the right-hand side. It is easily seen that 
$|-r + l\sin^2\phi| \geq |\sqrt{ r^2-(l\sin\phi)^2 }  \cos \phi|$ holds, since
\bqn
\displaystyle
|-r + l\sin^2\phi|^2 &-& |\sqrt{ r^2-(l\sin\phi)^2 } \cos \phi|^2 \nonumber \cr
& =& (r-l)^2 \sin^2\phi \geq 0.
\eqn 
Thus, due to Eq. (\ref{47}), 
$\frac{\partial^2 \Phi}{\partial l \partial r} \leq 0$
holds for $\phi \in (\pi,2\pi)$, which corresponds to Eq. (\ref{44}).

Next, using Eq. (\ref{46}) again, some lengthy calculations yield
\bqn\label{48}
\frac{\partial^2 \Phi}{\partial \phi \partial r}
& =& \frac{1}{2r\sin^2\phi}
 \Biggl[
  2l^2 \sin^3\phi \cos\phi - r^2\sin\phi \cos\phi \nonumber \\[5pt]
&& \hspace{10mm}+\sqrt{r^2 -(l\sin\phi)^2}(2\sin^2\phi-1)l \sin\phi \nonumber \\[5pt]
&& \hspace{10mm}+r^2\Bigl\{
        \phi-\pi +\arcsin (l\sin\phi/r) 
      \Bigr\}
 \Biggr].
\eqn
This time, we discuss the positivity of the right-hand side of Eq. (\ref{48}) for two cases with respect to the values of $\phi$: $(\phi,l,r) \in {\cal G}_1$ or ${\cal G}_2$.

The first case is when $(\phi,l,r) \in {\cal G}_1$, where $\cos \phi = -\sqrt{1-\sin^2 \phi}$ holds.
By taking into account $\phi-\pi= \arcsin(\sin(\phi-\pi)) = \arcsin(-\sin\phi)$,
denoting $y=\sin\phi$ leads to
\bqn\label{49}
\frac{\partial^2 \Phi}{\partial \phi \partial r}
 &=& (r^2y-2l^2 y^3) \sqrt{1-y^2} \nonumber \\[5pt]
 && \hspace{2mm} + ly(2y^2-1) \sqrt{r^2-(ly)^2} \nonumber \\[5pt]
 && \hspace{2mm} + r^2 \Bigl\{ \arcsin(ly/r) - \arcsin(y) \Bigr\},
\eqn
where $y \in (-1,0)$.
To show the positivity of the right-hand side of Eq. (\ref{49}), we regard it as a function of $r$
by fixing the variable $y$, which is denoted by $f(r)$. Then,
\bqn
g(r) \equiv \frac{f^\prime(r)}{2r} 
 &=& y \sqrt{1-y^2} \Bigl[ 1- \frac{l\sqrt{1-y^2}}{ \sqrt{ r^2-(ly)^2 }} \Bigr] \nonumber \\[5pt]
 && \hspace{2mm} + \Bigl\{ \arcsin(ly/r) - \arcsin(y) \Bigr\}. \nonumber
\eqn
Obviously, $g(l)=0$ holds, and in addition,
\bqn
&&g^\prime(r) = \frac{ly^3(l^2-r^2)}{r^2} \Bigl\{ r^2 - (ly)^2 \Bigr\}^{-3/2} \geq 0  \nonumber \\[5pt]
&& \hspace{30mm} \quad \forall r \geq 0, \; y \in (-1,0). \nonumber  
\eqn
This implies $g(r) \geq 0 \; \forall r \geq l$ for each $y \in (-1,0)$, and therefore,
$ f^\prime(r) \geq 0$ on the same interval. By noting that $f(l)=0$ holds due to Eq. (\ref{49}), 
we arrive at 
$$
\frac{\partial^2 \Phi}{\partial \phi \partial r} \geq 0  \quad \forall r \geq 0, \; y \in (-1,0).
$$ 

In the second case, when $(\phi,l,r) \in {\cal G}_2$, we first note that 
\bqn
 \phi-\pi &+& \arcsin(l\sin\phi/r) \nonumber \\[5pt]
 &=& \bigl( \phi- \frac{3\pi}{2} \bigr) + \frac{\pi}{2} + \arcsin(l\sin\phi/r) \nonumber \\[5pt]
 &=& \arccos(-\sin\phi) + \arccos(-l\sin\phi/r).\nonumber
\eqn
By applying this to Eq. (\ref{48}) and introducing $z=-\sin\phi \in (0,1),$ we obtain
\bqn
&&\frac{\partial^2 \Phi}{\partial \phi \partial r} = 
\frac{1}{2rz^2} 
 \Biggl[
  (r^2z-2l^2z^3) \sqrt{1-z^2} \nonumber \\[5pt]
&&\hspace{10mm}
  + rlz(1-2z^2) \sqrt{1-(lz/r)^2}
 \nonumber \\[5pt]
&& \hspace{10mm}
 + r^2 \Bigl\{
          \arccos(z) + \arccos(lz/r)
       \Bigr\}
 \Biggr]       
\eqn
In accordance with the definition of $\arccos(z),$ we have
\bqn
&&r^2 \arccos(z)+ (r^2z-2l^2z^3) \sqrt{1-z^2} \nonumber \\[5pt]
&& \hspace{5mm}= 2z( r^2-(lz)^2) \sqrt{1-z^2} + 2z^2 \int_z^1 \sqrt{1-t^2} \; {\rm d}t 
 \geq 0, \nonumber \\[5pt]
&& r^2 \arccos(lz/r)+ (rlz-2rlz^3) \sqrt{1-(lz/r)^2} \nonumber \\[5pt]
&& \hspace{5mm} =2rlz(1-z^2)\sqrt{1-(lz/r)^2} + 2r^2\int_{lz/r}^1\sqrt{1-t^2} \; {\rm d}t \nonumber \\[5pt]
&& \hspace{5mm} \geq 0.\nonumber
\eqn
Thus, we arrive at 
$$
\frac{\partial^2 \Phi}{\partial \phi \partial r} \geq 0  \quad \forall r \geq 0
$$ 
in this case also. This corresponds to Eq. (\ref{45}), and now, we have completed the proof of uniqueness by virtue of the 
preceding arguments.
\subsection{Proof of uniqueness regarding $l$}\label{unique_1p}
With $(\phi,r)$ provided, we solve the problem 
\bqn
&&g_4(l,\phi,r)=q
\eqn
with respect to $l$. By making use of Eq. (\ref{39}) in the preceding subsection,
we write 
\bqn\label{52}
 &&g_4(l,\phi,r) = lr\Biggl[1 + \frac{3}{4}\sqrt{1-\bigl( \gamma(\phi,l,r) \bigr)^2} \cos \phi \Biggr] \nonumber\\
 &&\hspace{11mm}
  -\frac 34 \bigl( l\sin \phi\bigr)^2 - \frac{r^2 \arcsin \gamma(\phi,l,r)}{4 \tan \phi} \nonumber \\
 && \hspace{10mm}
  + \frac{l^2 \zeta_1(\phi)}{2} \sin \phi \cos \phi 
  - \frac{(l\sin \phi)^2}{2} \log \Bigl( \frac{r}{l} \Bigr){\rm .} 
\eqn
If we regard the right-hand side of Eq. (\ref{52}) as a function of $l$ and denote it by $\Psi(l)$, then we have
\bqn\label{53}
 &&\Psi^\prime(l)
  = r + \sqrt{r^2- \bigl( l \sin \phi \bigr)^2 } \cos \phi  \nonumber \\
 &&\hspace{12mm}
   -l \Bigl( 1 + \log \Bigl( \frac{r}{l} \Bigr) \Bigr) \sin^2 \phi \nonumber  \\
 &&\hspace{12mm}
   + l \Bigl( \phi - \pi \arcsin  \gamma(\phi,l,r)  \Bigr) \sin \phi \cos \phi. 
\eqn
In the following, we show the non-negativity of $\Psi^\prime(l)$ for any $(\phi,l,r) \in {\cal G}_i \; (i=1,2)$ separately.

First, we consider the case $(\phi,l,r) \in {\cal G}_1$. 
In this case, we fix $l$ and $\phi$, and regard the right-hand of Eq. (\ref{53}) as a function of $r$, 
denoted by $F(r)$. Then, we show $F(r) \geq 0 \; \forall r\geq l$ with $(\phi,l,r) \in {\cal G}_1.$
By virtue of elementary calculations, it is easily seen that
\bqn
 F|_{r=l} &=& J_1 + J_2,\cr
J_1&\equiv&2l \sin^2\phi,\cr
J_2&\equiv&l\Bigl( \phi-\pi+\arcsin \Bigl( \frac{l\sin \phi}{r} \Bigr) \Bigr) \sin \phi \cos \phi. \nonumber 
\eqn
It is obvious that $J_1 \geq 0$ holds. Then, introducing $y=\sin\phi$,  we have $\phi -\pi = \arcsin (y)$ as we have seen in the previous subsection. This leads to
\bqn 
 &&\phi-\pi+\arcsin\Bigl( \frac{l\sin \phi}{r} \Bigr) = \arcsin(ly/r) - \arcsin(y) \geq 0,\nonumber
\eqn 
and we have $J_2\geq 0$ on ${\cal G}_1.$
These indicate
\begin{equation}
 F|_{r=l} \geq 0.\label{54}
\end{equation}
In addition, it is easily seen that
\bqn
 &&F^\prime(r) = 1+\frac{\sqrt{r^2-(l\sin \phi)^2} \cos \phi}{r} - \frac{l\sin^2 \phi}{r} \geq 0. \nonumber
\eqn
This, together with Eq. (\ref{54}), implies $\Psi^\prime(l) \geq 0$ on ${\cal G}_1.$

Next, we consider the case $(\phi,l,r) \in {\cal G}_2.$
Since $\phi \in (3\pi/2,2\pi)$ in this case, we have 
\bqn
 &&\phi -\pi + \arcsin \Bigl(\frac{l\sin\phi}{r} \Bigr)  \nonumber \\
 &&\hspace{10mm}
 = \arccos(-\sin\phi) + \arccos (-{l\sin\phi}/{r} ). \nonumber
\eqn
Therefore, by taking $z=-\sin\phi \in (0,1)$ and making use of $1+\log x \leq x \; \forall x \geq 0$, we have
\bqn
 &&\Psi^\prime(l) = r+ \sqrt{r^2-(lz)^2} \sqrt{1-z^2} - lz^2\Bigl( 1+\log \Bigl( \frac{r}{l}\Bigr) \Bigr) \nonumber\\
 &&\hspace{15mm}
  +lz\sqrt{1-z^2} \Bigl\{ \arccos(z) + \arccos (lz/r)\Bigr\} \nonumber \\
 &&\hspace{9mm}
  \geq (r+ lz^2)(1-z^2) \nonumber \\
 &&\hspace{15mm}
  + \Bigl( 1+ \frac{(lz)^2}{r^2} \Bigr) \sqrt{r^2-(lz)^2} \sqrt{1-z^2} \geq 0. \nonumber
\eqn
This completes the proof.
 

\section{Estimation method for $\{l_i,l_{i+1},\phi_i\}$}\label{concave_est_example}
Under limited cases such as a small number of concave vertices and large number of sensing results available, the following procedure theoretically can estimate $\{l_i,l_{i+1},\phi_i\}$ when the observability condition is satisfied. 
However, the estimated results are often unstable.

Let $\widehat{q(r)}$ be an estimator of $q(r)$.
Then, it is given as follows.
\bq
\widehat{q(r)}= 2\pi\sizex{\Omega}N(r)/n_s-2\hatlength r-2\pi\hatsize-2r^2\sum_i \widehat{g_1(\phi_i)}/8\label{q1}
\eq
where $\sum_i \widehat{g_1(\phi_i)}= 4\sum_{j=1,2}\{\pi\sizex{\Omega}N(s_j)/n_s-\hatlength s_j-\pi\hatsize\}/s_j^2$.

\begin{figure}[tb] 
\begin{center} 
\includegraphics[width=8cm,clip]{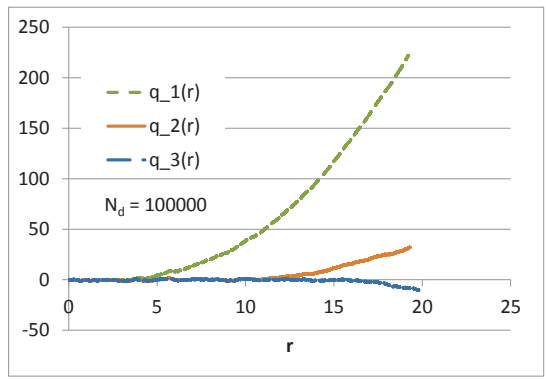} 
\caption{Example of $q_1(r),q_2(r)$} 
\label{concave_est} 
\end{center} 
\end{figure}

We expect that $q_1(r)\defeq\widehat{q(r)}\approx 0$ for $r<I_1$.
Therefore, by plotting $\widehat{q(r)}$, we can estimate where $I_1$ is and choose $s_3,s_4>I_1$ (Fig. \ref{concave_est}).
$s_3$ and $s_4$ should not be too large, because they should satisfy $s_3,s_4<I_2$.
Because $q(s_k)=-g_4(\phi_i,l_i,s_k)-s_k^2g_1(\phi_i)/8$ ($k=3,4$), we can estimate $\phi_i,l_i$ by using the calculated $q_1(s_3),q_1(s_4)$ and solving the following equations for $k=3,4$.
\bq
q_1(s_k)=-g_4(\phi_i,l_i,s_k)-s_k^2g_1(\phi_i)/8\label{appendix-1}
\eq

By using the estimated $\phi_i,l_i$, plot $q_{m+1}(r)\defeq q_m(r)+\sum_{j=3,4}(g_j(\widehat{\phi_i},\widehat{l_i},r)+r^2g_1(\widehat{\phi_i})/8)\bfone(C_{j}(\widehat{\phi_i},\widehat{l_i},r))$ with $m=1$.
We expect that $q_2(r)\approx 0$ for $r<I_2$.
Therefore, by plotting $q_2(r)$, we can estimate where $I_2$ is and choose $s_5,s_6>I_2$ (Fig. \ref{concave_est}).
For $I_2<s_5<s_6<I_3$, 
\bq
q_2(s_k)=-g_4(\phi_j,l_j,s_k)-s_k^2g_1(\phi_k)/8\label{appendix-2}
\eq
for $k=5,6$.
Hence, we can estimate $\phi_j,l_j$ by using the calculated $q_2(s_5), q_2(s_6)$ and solving the equations similar to Eq. (\ref{appendix-1}) for $k=5,6$.
If the estimated $\phi_j$ is approximately equal to $\phi_i$, judge that $j=i$ and adopt the estimated $l_j$ as $l_{i+1}$.

Repeat these steps by using $q_m(r)$.  We expect that we can estimate $\{l_i,l_{i+1},\phi_i\}_i$.


A numerical example is provided here to illustrate the estimation method mentioned above.
$T$ is that used in Subsection \ref{para-design}.
$T$ has one concave vertex with $\phi_1=4.89$ and $l_1=9.434, l_2=5.385$.
We performed a simulation with $N_d=100000$ and $\sizex{\Omega}=2500$.
According to Eq. (\ref{q1}), $q_1(r)=\widehat{q(r)}$ was obtained as a function of $r$ (Fig. \ref{concave_est}).
On the basis of this figure, we chose $s_3=6,s_4=8$.
Then, we obtained two equations (Eq. (\ref{appendix-1})) with $s_3=6,s_4=8$.
As a solution of these equations, we obtained an estimated $\phi_1=5.40$ and $l_1=2.69$.
By using these estimated results, we plotted $q_2(r)$ and chose $s_5=13,s_6=15$ (Fig. \ref{concave_est}).
We solved two equations (Eq. (\ref{appendix-2})) with $s_5=13,s_6=15$ and obtained the estimated $\phi_2=5.293$ and $l_2=10.79$.
On the basis of the plotted $q_3(r)$ (Fig. \ref{concave_est}), we judged that all $\{l_i,l_{i+1},\phi_i\}_i$ were estimated because $q_3(r)\approx 0$ for $0<\forall r<r_{max}$.
Because $\widehat{\phi_1}\approx\widehat{\phi_2}$, we judge that they are identical.

\section{Parameter design}\label{para-design}
With the proposed estimation method, we use parameters $r_{max}$ and $s_1,s_2$.
To obtain a good estimate, we should appropriately design the values of these parameters.
We use a simple quadrangle as $T$ and try various values of parameters in the estimation method to determine them.
The quadrangle has four edges with $l_1=9.434, l_2=5.385, l_3=10$, and $l_4=10\sqrt{2}$ and one concave vertex with a radian of $\phi_1=4.89$.

\begin{figure}[tb] 
\begin{center} 
\includegraphics[width=8cm,clip]{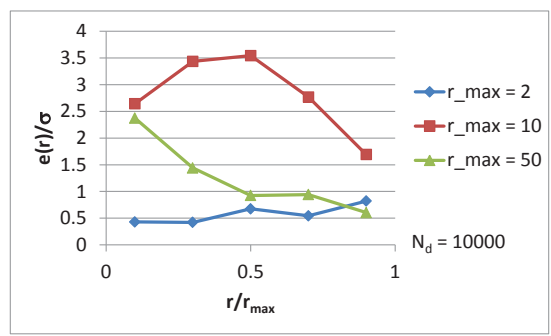} 
\caption{$e(r)/\sigma$ vs. $r_{max}$} 
\label{r_max} 
\end{center} 
\end{figure}

We determine $r_{max}$ by using $e(r)/\sigma$ for this quadrangle.
We conducted a simulation, and the results are shown in Fig. \ref{r_max} with $N_d\defeq N_0+N_+=10000$ for each simulation run.
In this figure, the average of $e(r)/\sigma$ of 20 runs is plotted against $r/r_{max}$ for each $r_{max}$.
According to this figure, $r_{max}\approx l_i$ or $r \approx l_i$ can detect non-convexity well if we judge it by evaluating that $e(r)/\sigma$ is larger than a threshold.
Specifically, $r \approx \min(l_i,l_{i+1})$ is suitable to detect non-convexity.

However, we do not know the exact $l_i$.
To be applicable to various $T$, it is better to use various $r$ to cover various $l_i$.
In the following, we use $r_{max}=20$ and $r/r_{max}=0.3,0.5,0.7$.
We can then obtain three $e(r)/\sigma$ corresponding to each value of $r/r_{max}$.
By using these three $e(r)/\sigma$, we choose the largest and judge the non-convexity of $T$ if it is larger than a threshold.

\begin{figure}[tb] 
\begin{center} 
\includegraphics[width=8cm,clip]{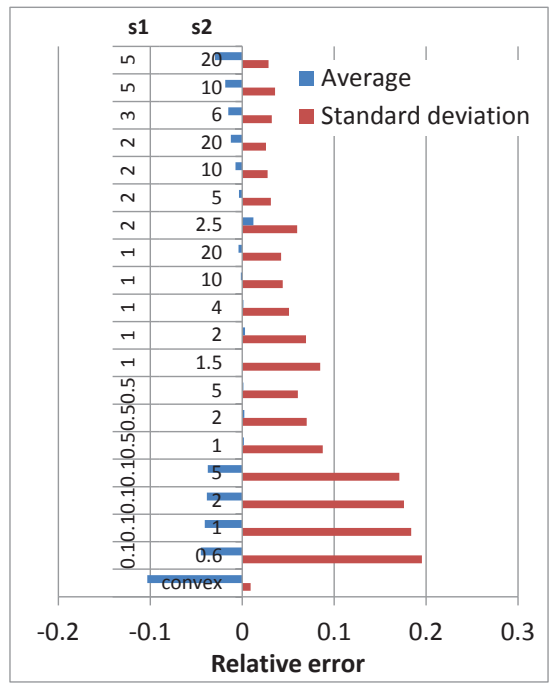} 
\caption{Perimeter estimation errors with various $s_1,s_2$} 
\label{perimeter} 
\end{center} 
\end{figure}

Next, we determined $s_1,s_2$ by using the simulation results for this quadrangle with $N_d=10000$.
The results are shown in Fig. \ref{perimeter}.
For various $s_1$ and $s_2$, $\lengthx{T}$ is estimated with the proposed method using Eqs. (\ref{length}) and (\ref{peri_non-convex}), and its relative error is evaluated.
The last one is the result assuming the convexity and calculated using Eq. (\ref{length}).
As shown in this figure, $\hatlength$ assuming the convexity has a large negative bias, that is, underestimates $\lengthx{T}$ for a non-convex $T$.
However, the relative error has a very small variance.
In comparison, $\lengthx{T}$ estimated with the proposed method has a very small bias and a large variance.
Therefore, the estimation of $\lengthx{T}$ for a non-convex $T$ requires a large number of samples.
In general, (i) a too small or too large $s_1$ may cause a bias, and (ii) a large $s_2-s_1$ has a small variance of the relative error.
As a result, $s_1=1$ and $s_2\approx l_i,l_{i+1}$ are appropriate combinations.
Because we do not know the exact $l_i$, we calculate $\hatlength$ using Eq. (\ref{length}) with a small number of samples and use, for example, $s_2=\hatlength/10$.

\end{document}